\documentclass[submission, Proceedings]{SciPost}

\hypersetup{
    colorlinks,
    linkcolor={red!50!black},
    citecolor={blue!50!black},
    urlcolor={blue!80!black}
}

\usepackage[bitstream-charter]{mathdesign}
\DeclareSymbolFont{usualmathcal}{OMS}{cmsy}{m}{n}
\DeclareSymbolFontAlphabet{\mathcal}{usualmathcal}

\usepackage{caption}
\usepackage{subcaption}

\def\ptv{P_{\rm{T}}}
\def\gevc{\rm{GeV}/c}
\def\gevc2{\rm{GeV}/c^2}
\def\Mhh{M_{\rm{hh}}}
\def\pht{P_{\rm{T}}}

\def\Acoll{A_{\rm{UT}}^{\sin(\phih+\phi_{\rm S}-\pi)}}
\def\Asiv{A_{\rm{UT}}^{\sin(\phih-\phi_{\rm S})}}

\def\phih{\phi_{\rm{hh}}}

\def\acoll{a_{\rm{UT}}^{\sin(\phih+\phi_{\rm S}-\pi)}}
\def\asiv{a_{\rm{UT}}^{\sin(\phih-\phi_{\rm S})}}

\def\Acollbg{A_{\rm{UT},bg}^{\sin(\phih+\phi_{\rm S}-\pi)}}

\def\rhoz{$\rho^0$}
\def\aUTX{a_{\rm{UT}}^{\sin(\phi_{\rm X})}}
\def\AUTbgX{A_{\rm{UT},bg}^{\sin(\phi_{\rm X})}}
\def\AUTX{A_{\rm{UT}}^{\sin(\phi_{\rm X})}}
\def\xu{\hat{\textbf{x}}}
\def\zu{\hat{\textbf{z}}}
\begin{document}

\begin{center}{\Large \textbf{
Transverse spin asymmetries for inclusive $\rho^0$ production in SIDIS at COMPASS\\
}}\end{center}

\begin{center}
A. Kerbizi\textsuperscript{$\star$} \\
for the COMPASS Collaboration
\end{center}

\begin{center}
Trieste Section of INFN, University of Trieste, \\
Dept. of Physics, 34127 Trieste, Italy
\\
* albi.kerbizi@ts.infn.it
\end{center}



\definecolor{palegray}{gray}{0.95}
\begin{center}
\colorbox{palegray}{
  \begin{tabular}{rr}
  \begin{minipage}{0.1\textwidth}
    \includegraphics[width=22mm]{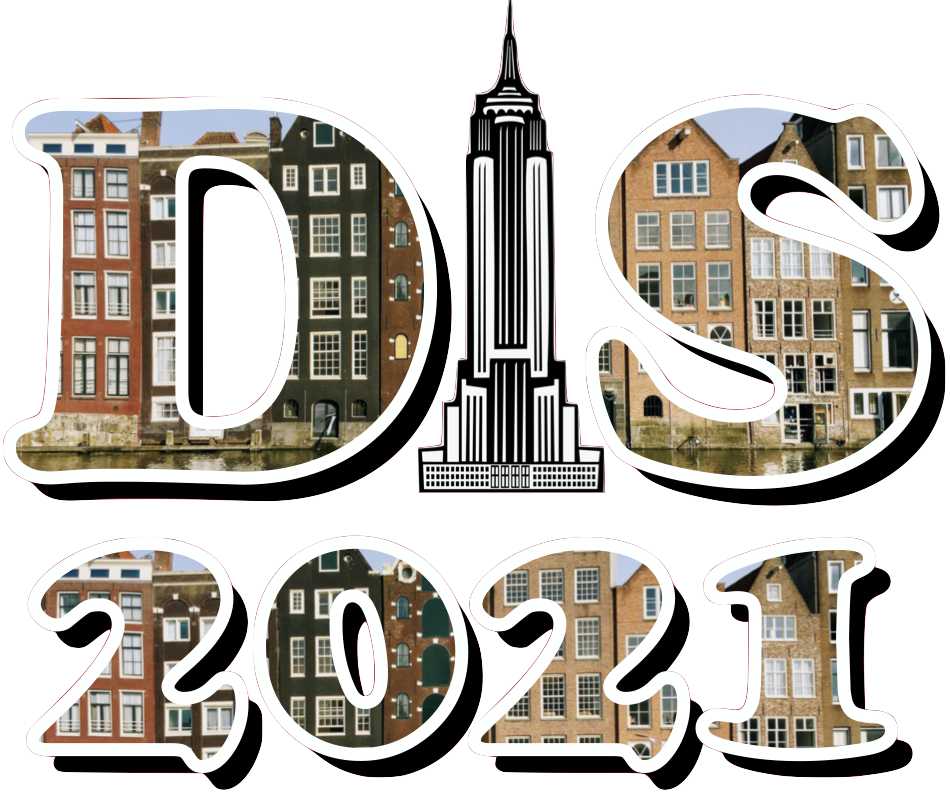}
  \end{minipage}
  &
  \begin{minipage}{0.75\textwidth}
    \begin{center}
    {\it Proceedings for the XXVIII International Workshop\\ on Deep-Inelastic Scattering and
Related Subjects,}\\
    {\it Stony Brook University, New York, USA, 12-16 April 2021} \\
    \doi{10.21468/SciPostPhysProc.?}\\
    \end{center}
  \end{minipage}
\end{tabular}
}
\end{center}

\section*{Abstract}
{\bf
The production of vector mesons in SIDIS is a particularly interesting channel to study the polarized fragmentation and related phenomena. Preliminary COMPASS results for the first measurement of inclusive $\rho^0$ Collins and Sivers transverse-spin asymmetries are presented here for the first time.
The analysis is based on the data-set collected by COMPASS in $\textbf{2010}$ using a $\textbf{160}\, \boldsymbol{\rm{GeV}/c}$ $\boldsymbol{\mu^+}$ beam and a transversely polarized $\boldsymbol{NH_3}$ target.
The asymmetries are extracted as function of Bjorken-$\boldsymbol{x}$, of the total transverse momentum of the oppositely charged hadron pair and of the fraction of available energy $\boldsymbol{z}$ carried by the pair. Indications for positive Collins and Sivers asymmetries are obtained as expected from models.
}


\section{Introduction}
\label{sec:intro}
Transverse spin asymmetries (TSAs) in Semi-Inclusive Deep Inelastic Scattering (SIDIS) like the Collins, dihadron and Sivers asymmetries are being measured since 2005. TSAs have been measured in SIDIS off proton, deuteron or neutron targets for unidentified charged hadrons or for identified pions, kaons and protons by the HERMES \cite{hermes-ssa}, COMPASS \cite{COMPASS-collins-sivers,compass-dihadron} and JLab \cite{jlab-ssa} collaborations. Valuable information on the nucleon structure encoded in parton distribution functions (PDFs) and on the fragmentation process encoded in fragmentation functions (FFs) have been extracted from such asymmetries. The most remarkable results concern the extraction of the transversity and the Sivers PDFs that turned out to be different from zero. Similar results for the transversity function have been obtained from the measurement of the dihadron asymmetries \cite{compass-dihadron,hermes-dihadron}. 

An other interesting and unexplored class of TSAs are those related to the inclusive production of vector mesons which probe a new and unknown class of FFs \cite{Bacchetta:Spin1}. For instance, models suggest that the Collins effect for the production of $\rho$ mesons is opposite and smaller by a factor of three with respect to that of positive pions \cite{Czyzewski-vm,Kerbizi:PhD}.
The measurement of TSAs for vector mesons is challenging due to the lower statistics than pseudoscalar mesons and the high combinatorial background. Although their knowledge is crucial to shed light on the spin-dependence of the fragmentation process, they had never been measured.
In this article we present preliminary results of the first ever measurement of the Collins and Sivers asymmetries for inclusive production of \rhoz{} in SIDIS off transversely polarized protons performed recently by the COMPASS Collaboration.

The data sample used for this analysis is introduced in Sec. \ref{sec:data sample}. The procedure followed for the extraction of TSAs is explained in Sec. \ref{sec:asymmetry extraction procedure} and the final \rhoz{} TSAs are presented in Sec. \ref{sec:TSA in Mhh regions}. The conclusions are given in Sec. \ref{sec:conclusions}.

\section{The data sample}\label{sec:data sample}
We have analyzed the full SIDIS data sample collected in 2010 by scattering a $\mu^+$ beam with $160\,\rm{GeV}/c$ momentum off a transversely polarized $NH_{3}$ target, already used for many published results \cite{COMPASS-collins-sivers,compass-dihadron}. We require the virtuality of the exchanged photon to be $Q^2 > 1.0\, (\rm{GeV}/c)^2$ and the invariant mass of the final hadronic system $W > 5\, \rm{GeV}/c^2$. Also the Bjorken $x$ variable is taken $0.003 < x < 0.7$ and the fraction of the photon energy carried by the scattered muon $0.1 < y < 0.9$. On the hadron side, for each hadron the fractional energy is required to be larger than $0.1$ and the transverse momentum with respect to the exchanged photon direction larger than $0.1 \,\rm{GeV}/c$.

Then, all the pairs $h_1h_2 = h^+h^-, h^+h^+, h^-h^-$ in the same event have been selected as follows. The cut on missing energy  $E_{\rm{miss}} > 3.0\,\rm{GeV}$ has been applied to reject the exclusive $h^+h^-$ pairs. The missing energy is defined to be $E_{\rm{miss}}=(M_{\rm X}^2-M_{\rm p}^2)/2M_{\rm p}$, where $M_{\rm X}^2=(q+P_{\rm p}-P)^2$ with $q$, $P_{\rm p}$ and $P$ the four-momenta respectively of the exchanged photon, of the target proton and of the hadron pair, and $M_{\rm p}$ is the proton mass. The fractional energy of the pair $z=z_{1}+z_{2}$ is required to be larger than $0.3$, in order to enhance the fraction of \rhoz{} mesons. The additional cuts $z<0.95$, $0.1\,\rm{GeV}/c < \pht < 4.0\,\rm{GeV}/c$
with $\vec{P}_{\rm T}=\vec{P}_{1\rm T}+\vec{P}_{2\rm T}$ being the transverse momentum of the pair, and $0.35\,\gevc2 < \Mhh < 3.0\,\gevc2$  have been applied.
 
%
\begin{figure}[!h]
    \vspace{-1em}
     \centering
     \begin{subfigure}[b]{0.58\textwidth}
         \centering
         \includegraphics[width=1.0\textwidth]{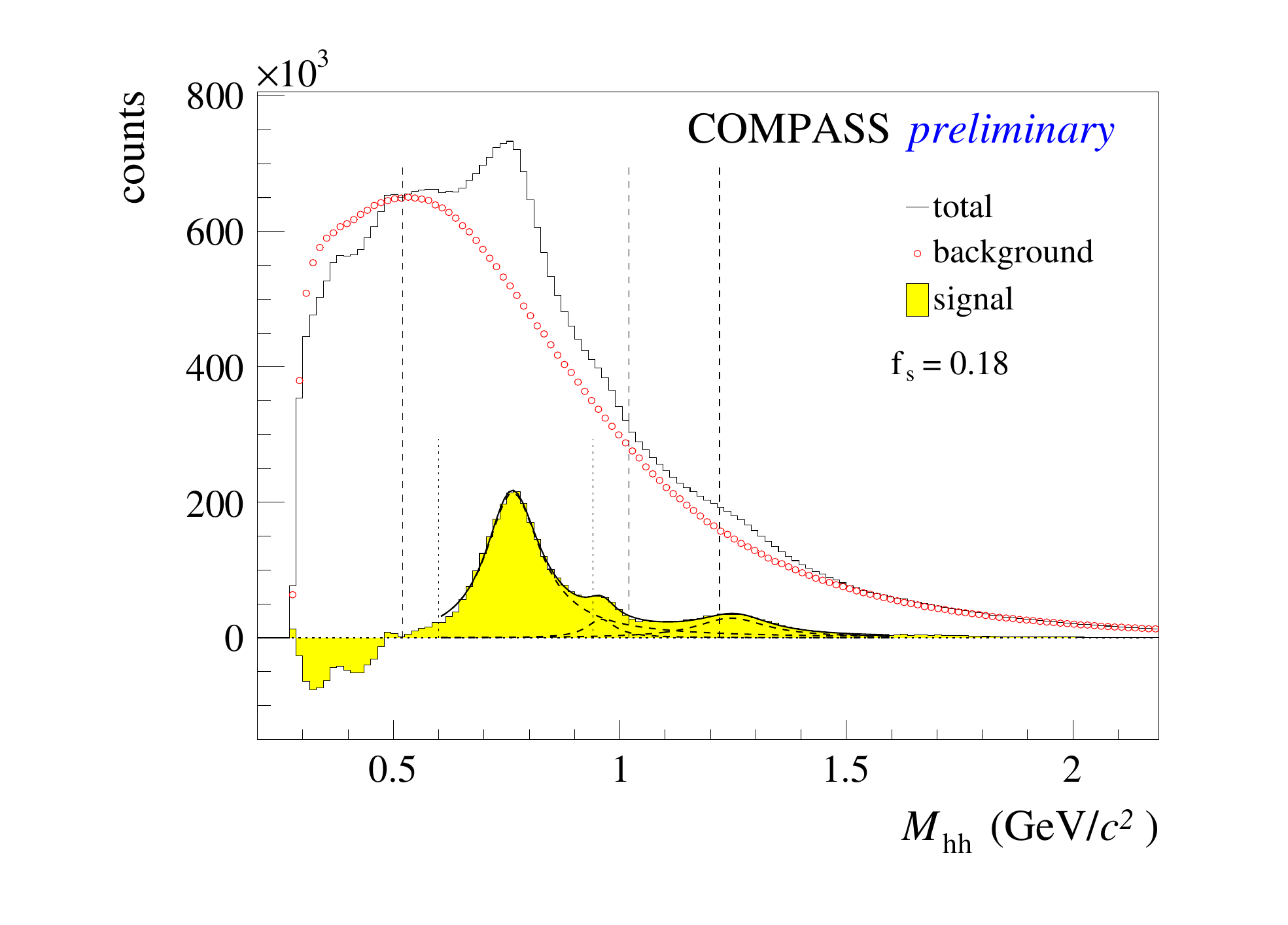}
     \end{subfigure}
     \vspace{-1.0em}
    \caption{Total invariant mass distribution of $h^+h^-$ pairs (continuous empty histogram), estimated background from the $h^+h^+$ and $h^-h^-$ pairs (points) and the signal distribution (filled histogram).}
    \label{fig:sig extraction}
\end{figure}
The final invariant mass distribution of $h^+h^-$ pairs is shown by the continuous empty histogram in Fig. \ref{fig:sig extraction}.
The $\rho^0(770)$ peak is clearly visible as well as the structures corresponding to the $f_0(980)$ and $f_2(1240)$ mesons. As it can be seen the combinatorial background under the $\rho^0(770)$ peak is large.
The points show the normalized invariant mass distribution of $h^+h^+$ and $h^-h^-$ pairs, that we take as background distribution.
This recipe gives a good estimation of the background in the signal region, although the agreement is not good at low invariant mass. Also, the systematic uncertainty in the background estimation has been included in the systematic uncertainty of the final \rhoz{} asymmetries. The filled histogram shows the background subtracted yield of $h^+h^-$ pairs.

\section{Procedure for the extraction of TSAs}\label{sec:asymmetry extraction procedure}
We divided the invariant mass range in four regions, defined as region $I$ for $\Mhh\in[0.35,0.52]$ $\gevc2$, region $II$ for $\Mhh\in[0.60,0.94]$ $\gevc2$, region $III$ for $\Mhh\in[1.02,1.22]$ $\gevc2$ and region $IV$ for $\Mhh\in[1.22,3.0]\,\rm{GeV}/c^2$,
as shown by the vertical lines in Fig. \ref{fig:sig extraction}. The region $II$ is the "\rhoz{} region" which contains the \rhoz{} peak, whereas regions $I$ and $III$ are the "side band" regions given by the combinatorial background.

The strategy applied for the extraction of the \rhoz{} TSAs consists in the following steps:
\begin{itemize}
    \itemsep0em
    \item[a.] Evaluate the \rhoz{} fraction $f_{s}$ in the \rhoz{} region.
    \item[b.] Measure the transverse spin asymmetry $\aUTX$ of the $h^+h^-$ pairs in the \rhoz{} region.
    \item[c.] Measure the background transverse spin asymmetry $\AUTbgX{}$ from the side band regions.
    \item[d.] Subtract the background transverse spin asymmetry to obtain the final asymmetry by using
    \begin{equation}\label{eq:subtraction}
        \AUTX = \left[\aUTX-(1-f_s)\,\AUTbgX\right]\times f_s^{-1}.
    \end{equation}
\end{itemize}
The azimuthal angle $\phi_{\rm X}$ indicates either the Collins angle $\phi_{Coll}=\phih+\phi_{\rm S}-\pi$ or the Sivers angle $\phi_{Siv}=\phih-\phi_{\rm S}$. The angle $\phih$ is the azimuthal angle of the transverse momentum $\vec{P}_{\rm T}$ of the $h^+h^-$ pair whereas $\phi_{\rm S}$ is the azimuthal angle of the target transverse polarization. Both angles are defined in the so called GNS frame where the exchanged photon and the target nucleon are collinear with respect to each other and the $\xu-\zu$ plane is the lepton scattering plane. The $\zu$ axis is taken along the exchanged photon momentum and the $\xu$ axis along the transverse momentum of the scattered muon.

The transverse spin asymmetries have been extracted in six bins of $x$, $z$ and $\ptv$ by using standard COMPASS methods \cite{COMPASS-2006}. 

\begin{figure}[b]
     \vspace{-1.0em}
     \centering
     \begin{subfigure}[b]{0.65\textwidth}
         \centering
         \includegraphics[width=1.0\textwidth]{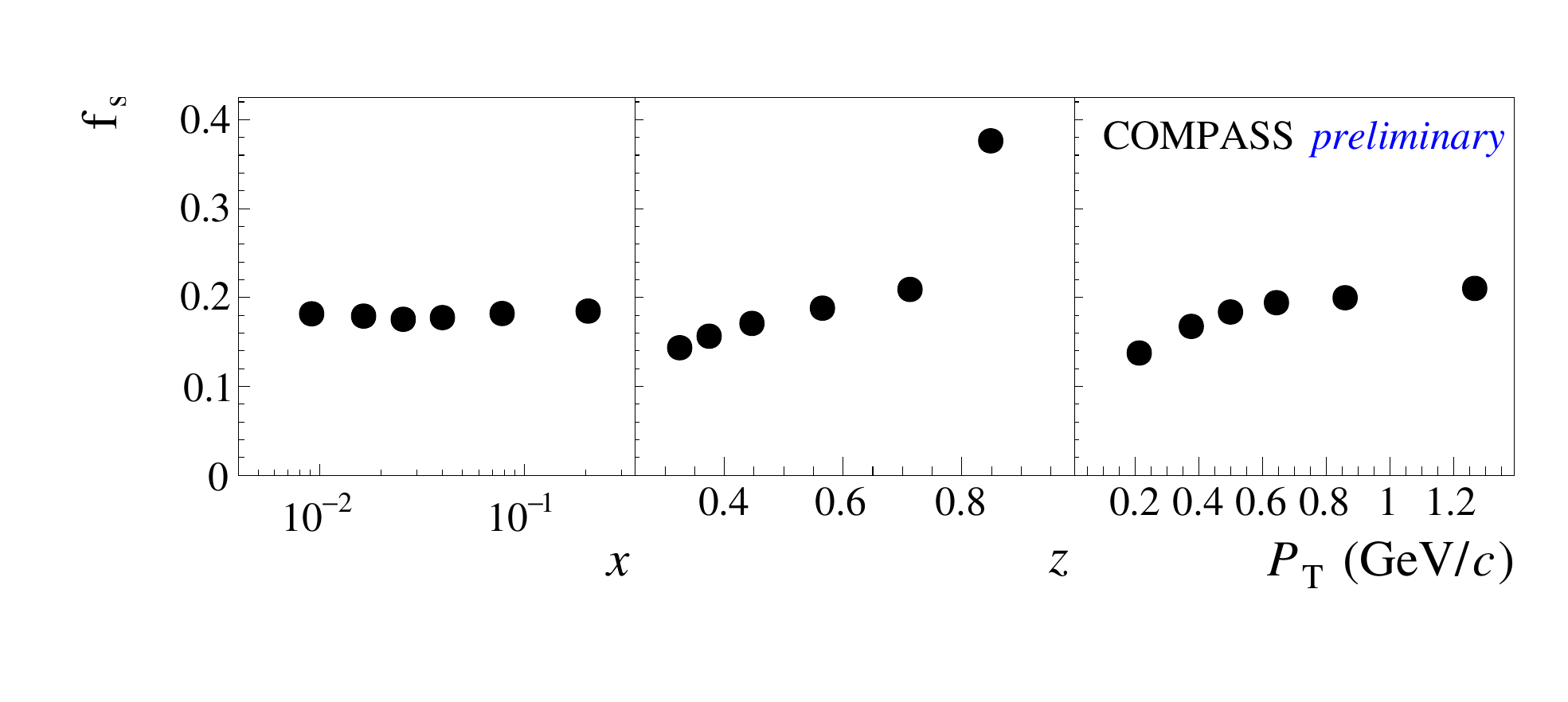}
     \end{subfigure}
     \vspace{-1.2em}
    \caption{The \rhoz{} signal fraction as function of $x$ (left plot), $z$ (middle plot) and $\ptv$ (right plot).}
    \label{fig:sig fraction}
\end{figure}

To measure the fraction of \rhoz{} mesons in the invariant mass region $II$, the background distribution, shown by the points in Fig. \ref{fig:sig extraction}, is subtracted from the total $h^+h^-$ distribution. 
The resulting signal distribution is shown by the filled histogram in Fig. \ref{fig:sig extraction} where the peaks corresponding to $\rho^0(770)$, $f_0(980)$ and $f_2(1270)$ mesons are clearly visible. A fit of the signal distribution with the sum of three Breit-Wigner functions shown in the figure demonstrates that the subtraction procedure is clean.
Finally, the fraction $f_s$ of \rhoz{} mesons in region $II$ is calculated dividing the \rhoz{} counts by the total $h^+h^-$ pairs in the same region.

This procedure has been applied to all $x$, $z$ and $\ptv$ bins and the values of $f_s$ are shown in Fig. \ref{fig:sig fraction}. We find $f_s$ to be almost constant and about $0.18$ as function of $x$ and it increases with $\ptv$ and $z$. In particular the high value (about $0.38$) in the last $z$ bin can be understood in terms of the string fragmentation model \cite{Lund1983}.

\begin{figure}[!t]
    \vspace{-2em}
     \centering
     \begin{subfigure}[b]{0.6\textwidth}
         \centering
         \includegraphics[width=1.0\textwidth]{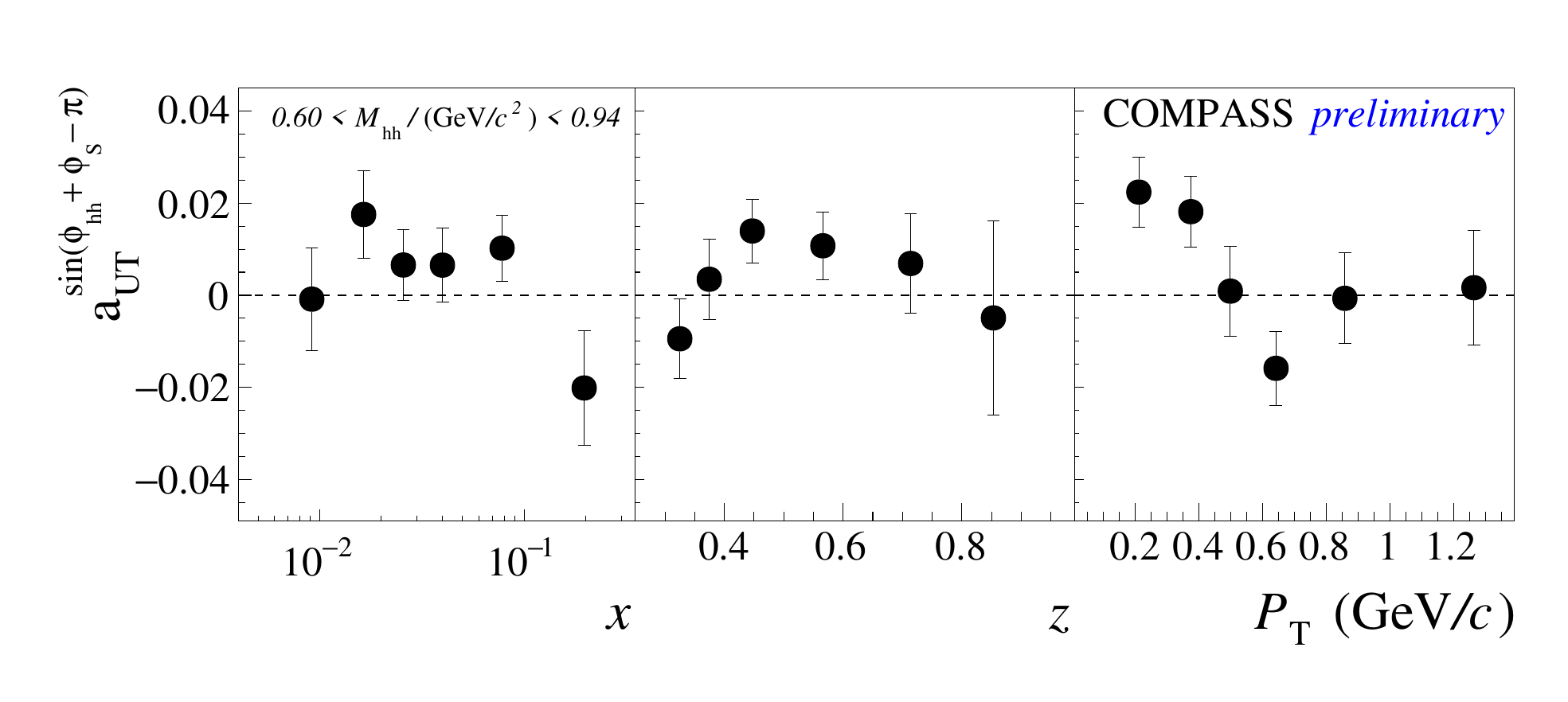}
     \end{subfigure}\\
     \vspace{-0.8em}
     \begin{subfigure}[b]{0.6\textwidth}
         \centering
         \includegraphics[width=1.0\textwidth]{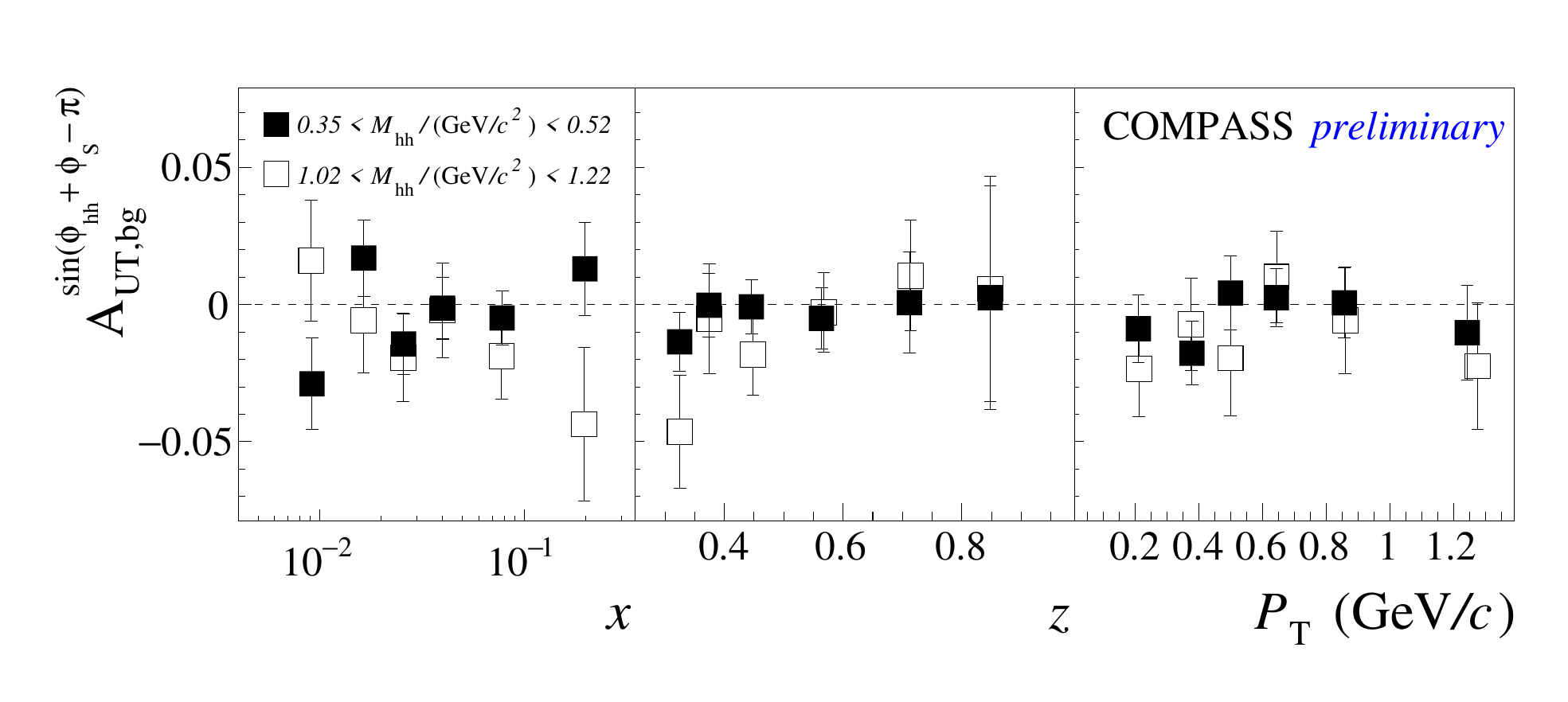}
     \end{subfigure}\\
     \vspace{-0.8em}
    \hspace{1em}
     \begin{subfigure}[b]{0.58\textwidth}
         \centering
         \includegraphics[width=1.0\textwidth]{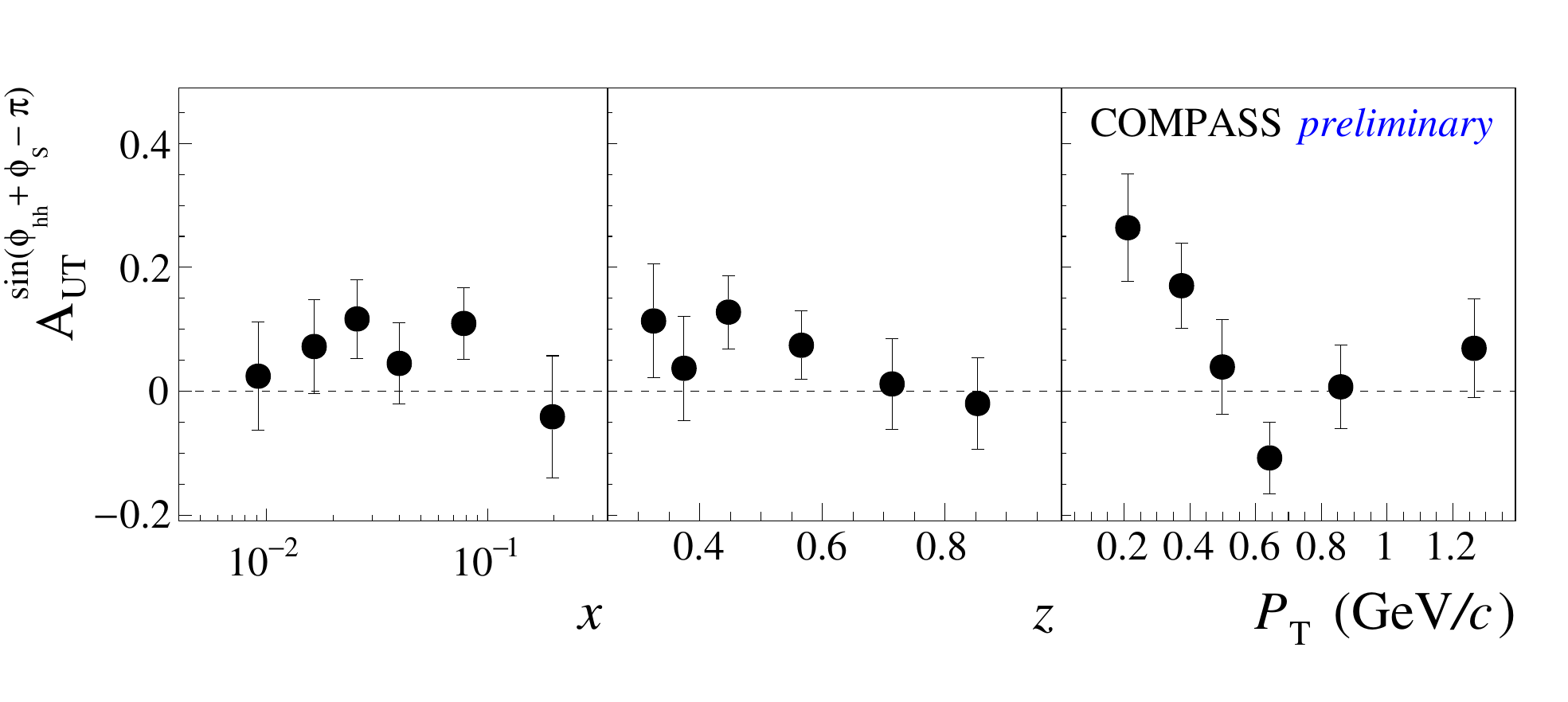}
         \label{fig:y equals x}
     \end{subfigure}\vspace{-1.2em}
        \caption{Upper panel: Collins asymmetry in region $II$ as function of $x$ (left plot), $z$ (middle plot) and $\ptv$ (right plot). Middle panel: corresponding asymmetries in region $I$ (full points) and $II$ (open points). Lower panel: final Collins asymmetry for \rhoz{} mesons. The errorbars show the statistic uncertainties only. The systematic uncertainties for the \rhoz{} asymmetry have been estimated to be $\sigma_{sys}=0.3\sigma_{stat}$.}
        \label{fig:collins asymmetry}
\end{figure}

\section{TSAs for \rhoz{} mesons}\label{sec:TSA in Mhh regions}
The Collins asymmetry $\acoll{}$ in the \rhoz{} region is shown in the upper panel of Fig. \ref{fig:collins asymmetry} as function of $x$, $z$ and $\ptv$. In spite of the large statistical uncertainties, the asymmetry is positive and in particular at low $\ptv$ and for intermediate $z$ values. The corresponding asymmetries in the side band regions $I$ and $III$ are shown in the middle panel of the same figure. They are similar and compatible with zero.

The final Collins asymmetry $\Acoll{}$ for \rhoz{} mesons is shown in the lower panel of Fig. \ref{fig:collins asymmetry}. The uncertainties are statistical only. It has been obtained using in each bin the background asymmetry $\Acollbg{}$ according to Eq. (\ref{eq:subtraction}). The background asymmetry is calculated as the arithmetic average of the asymmetries in regions $I$ and $III$. In spite of the large uncertainties, the \rhoz{} Collins asymmetry is positive in all kinematic bins and a clear effect can be seen for $\ptv<0.5\,\rm{GeV}/c$.
This is consistent with the expectations from polarized quark fragmentation models \cite{Czyzewski-vm,Kerbizi:PhD}.

\begin{figure}[!t]
    \vspace{-2em}
     \centering
     \begin{subfigure}[b]{0.6\textwidth}
         \centering
         \includegraphics[width=1.0\textwidth]{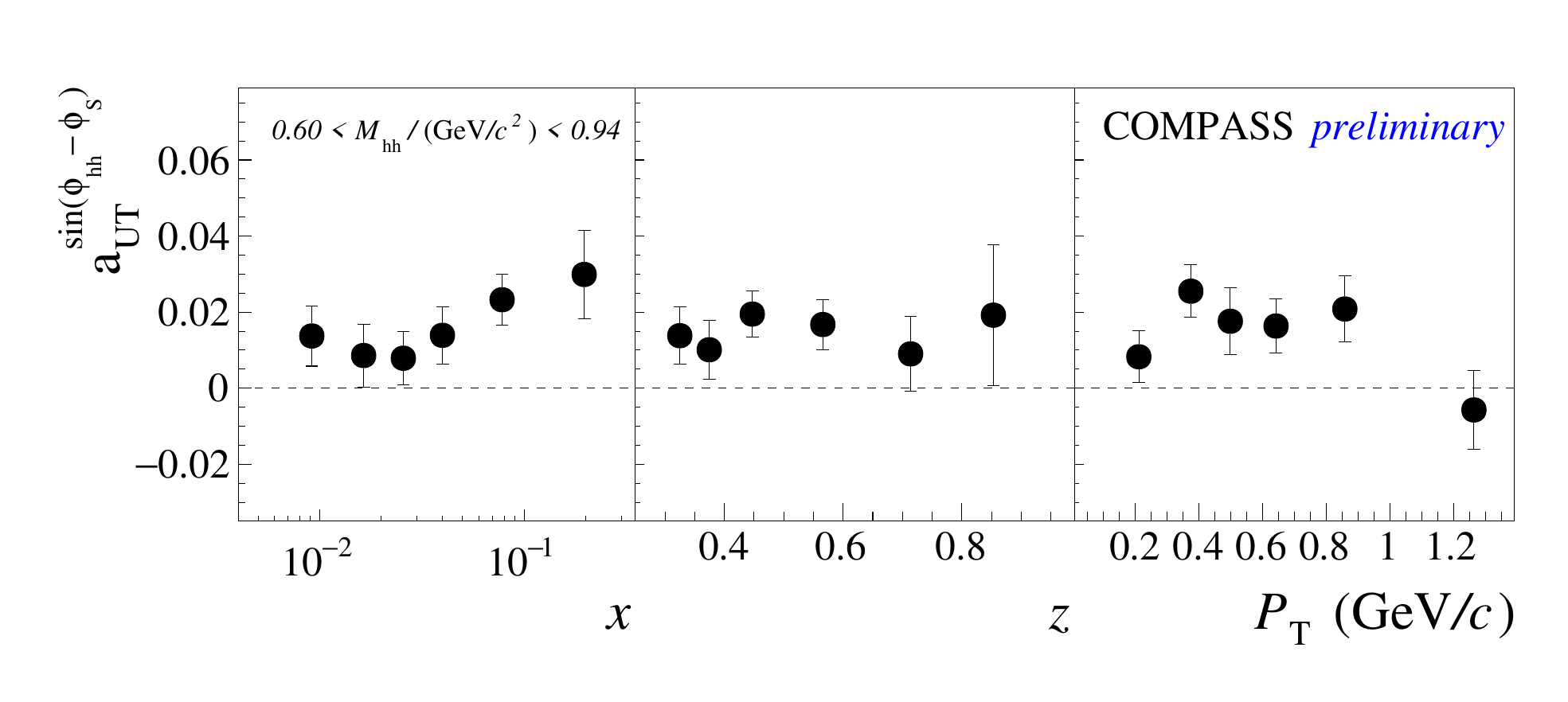}
     \end{subfigure}\\
     \vspace{-0.8em}
     \begin{subfigure}[b]{0.6\textwidth}
         \centering
         \includegraphics[width=1.0\textwidth]{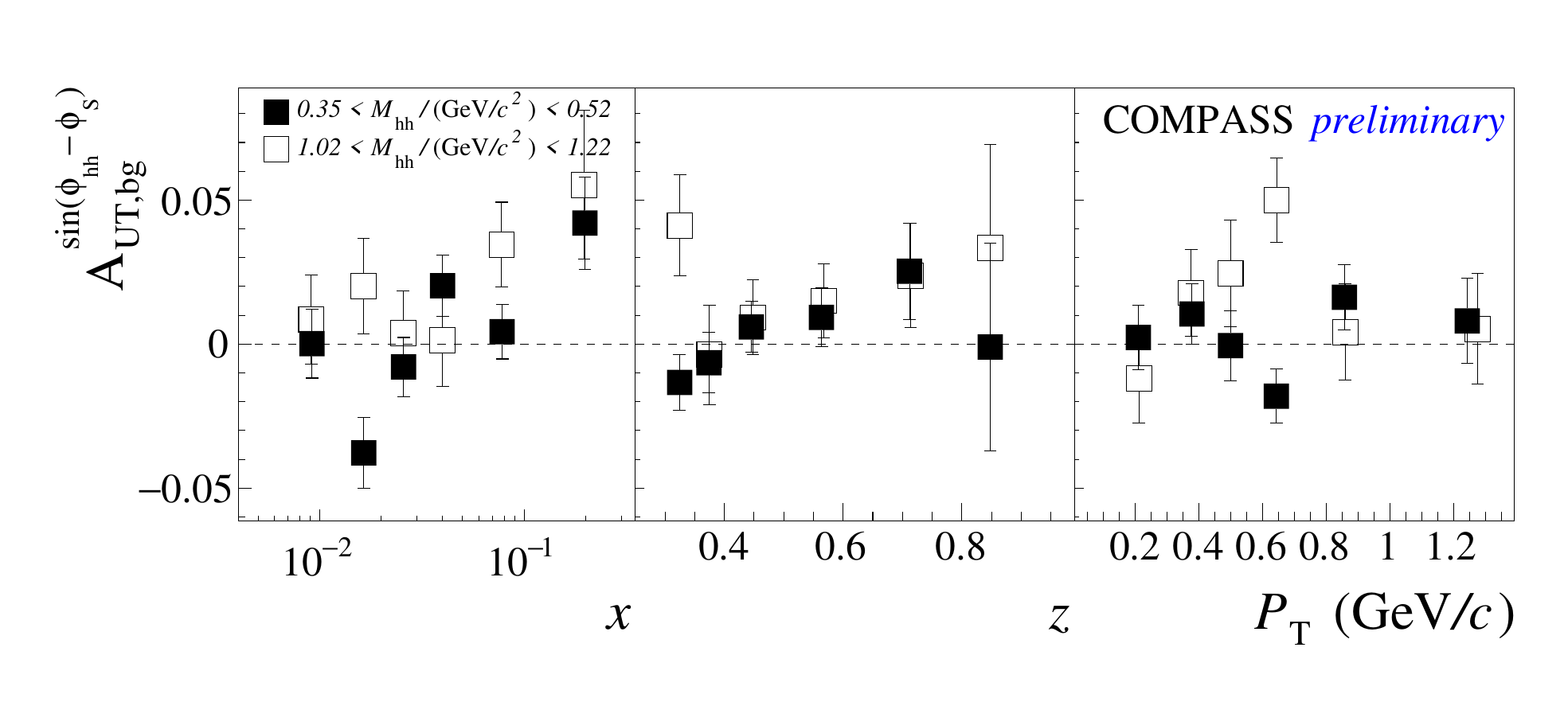}
     \end{subfigure}\\
     \vspace{-0.8em}
    \hspace{1em}
     \begin{subfigure}[b]{0.58\textwidth}
         \centering
         \includegraphics[width=1.0\textwidth]{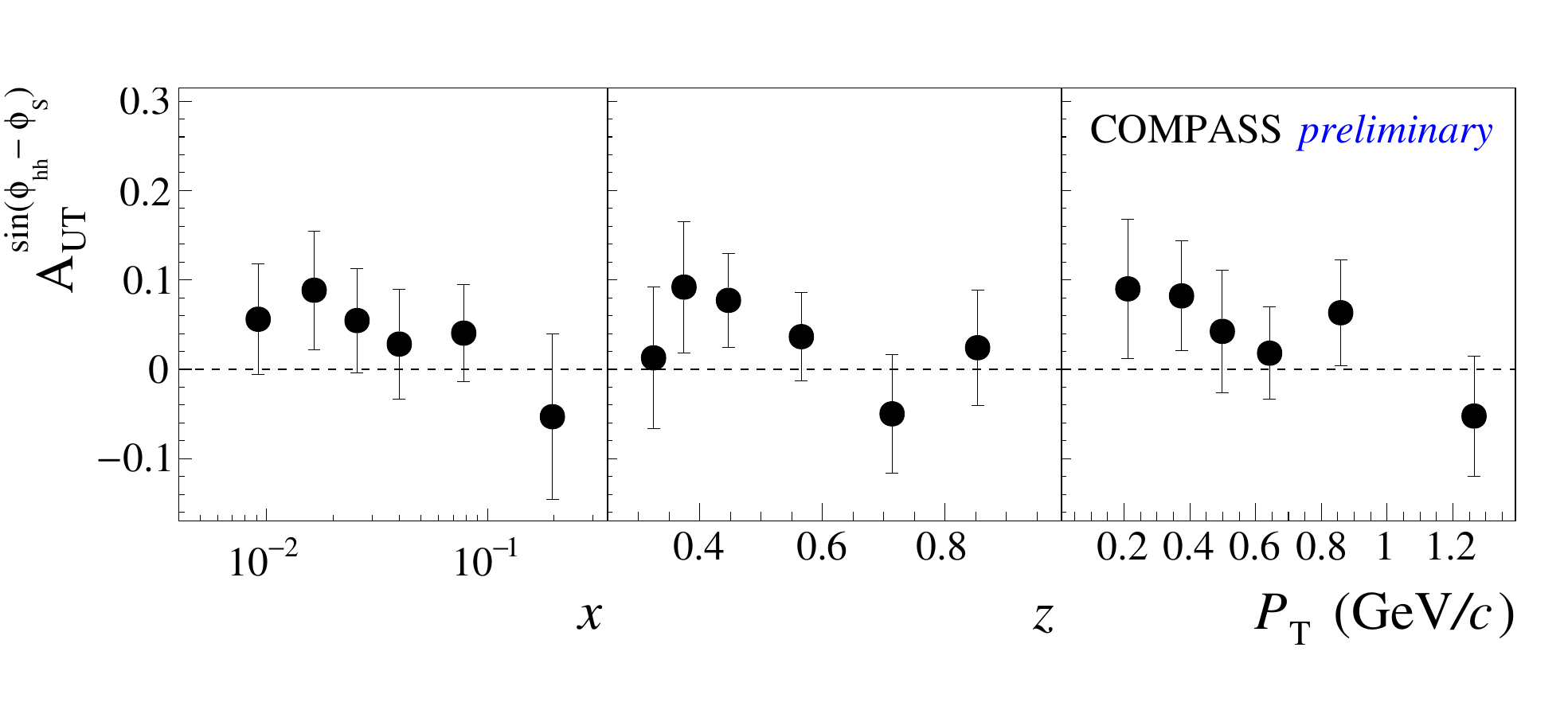}
         \label{fig:y equals x}
     \end{subfigure}\vspace{-1.2em}
        \caption{Upper panel: Sivers asymmetry in region $II$ as function of $x$ (left plot), $z$ (middle plot) and $\ptv$ (right plot). Middle panel: corresponding asymmetries in region $I$ (full points) and $II$ (open points). Lower panel: final Sivers asymmetry for \rhoz{} mesons. Statistic uncertainties only. Systematic uncertainties for the \rhoz{} asymmetry have been estimated to be $\sigma_{sys}=0.3\sigma_{stat}$.}
        \label{fig:Sivers asymmetry}
\end{figure}

The Sivers asymmetries are shown in Fig. \ref{fig:Sivers asymmetry}. The uncertainties are again statistical only. The asymmetry $\asiv{}$ in the \rhoz{} region is large and positive but also the 
background asymmetry in the side band regions is large, as can be seen from Fig. \ref{fig:Sivers asymmetry}.
The final Sivers asymmetry for \rhoz{} mesons $\Asiv{}$ is shown in the bottom panel in Fig. \ref{fig:Sivers asymmetry}. Also in this case we find a hint for a positive asymmetry, as can be expected from considerations based on the parton model and the measurements of the Sivers asymmetries for unpolarized hadrons \cite{COMPASS-collins-sivers, hermes-ssa}, compatible with the side bands asymmetries.

\section{Conclusion}\label{sec:conclusions}
The COMPASS Collaboration has performed the first measurement of the Collins and Sivers transverse spin asymmetries for \rhoz{} mesons inclusively produced in SIDIS off transversely polarized protons. In spite of the low statistics, an indication for a positive Collins asymmetry is found, as expected from models of the polarized fragmentation process. The result provides relevant information for the tuning of the free parameters of models. Also an indication for a positive Sivers asymmetry is found, in agreement with the parton model. Most important, this work shows the feasibility of the measurement of TSAs for inclusive vector meson production which could be done with much higher precision at future facilities.




\begin{appendix}


\end{appendix}



\bibliography{SciPost_LaTeX_Template.bbl}

\nolinenumbers

\end{document}